\documentclass[12pt,a4paper]{article}
\usepackage{graphicx}
\usepackage{times}
\title{\bf Circumstellar Material Around Evolved Massive Stars}

\author{Nathan Smith$^{1,2}$\\
  \vspace{1cm}\\
  \normalsize $^1$ Steward Observatory, University of Arizona, 933
  N. Cherry Ave., Tucson, AZ, 85721, USA \\
  \normalsize $^2$ Astronomy Department, University of California, 601
  Campbell Hall, Berkeley, CA 94720, USA}
\date{\mbox{}}
\begin{document}
\maketitle
\pagestyle{empty}
%
%
\def\bull{\vrule height .9ex width .8ex depth -.1ex}
\makeatletter
\def\ps@plain{\let\@mkboth\gobbletwo
\def\@oddhead{}\def\@oddfoot{\hfil\tiny\bull\quad
``The multi-wavelength view of hot, massive stars''; 39$^{\rm th}$ Li\`ege Int.\ Astroph.\ Coll., 12-16 July 2010 \quad\bull}%
\def\@evenhead{}\let\@evenfoot\@oddfoot}
\makeatother
%
%
\def\beginrefer{\section*{References}%
\begin{quotation}\mbox{}\par}
\def\refer#1\par{{\setlength{\parindent}{-\leftmargin}\indent#1\par}}
\def\endrefer{\end{quotation}}
%
{\noindent\small{\bf Abstract:} 

  I review multiwavelength observations of material seen around
  different types of evolved massive stars (i.e. red supergiants,
  yellow hypergiants, luminous blue variables, B[e] supergiants, and
  Wolf-Rayet stars), concentrating on diagnostics of mass,
  composition, and kinetic energy in both local and distant examples.
  Circumstellar material has significant implications for the
  evolutionary state of the star, the role of episodic mass loss in
  stellar evolution, and the roles of binarity and rotation in shaping
  the ejecta.  This mass loss determines the type of supernova that
  results via the stripping of the star's outer layers, but the
  circumstellar gas can also profoundly influence the immediate
  pre-supernova environment. Dense circumstellar material can actually
  change the type of supernova that is seen when it is illuminated by
  the supernova or heated by the blast wave.  As such, unresolved
  circumstellar material illuminated by distant supernovae can provide
  a way to study mass loss in massive stars in distant environments.

}

\section{Introduction and Scope}

There is a great variety of circumstellar material around massive
stars.  In the interest of brevity, I will restrict myself to
discussing stars with detectable circumstellar {\it nebulae}, as
distinguished from steady stellar {\it winds} or {\it disks} where the
primary observable signature is seen in the spectrum and generally
arises within several stellar radii (this was discussed in detail
during the first session of this conference; see reviews by Owocki and
Martins).  By virtue of their high luminosity and radiation-driven
winds, all massive stars shed mass in stellar winds, except perhaps at
near zero metallicity.  However, not all stars are surrounded by
detectable circumstellar nebulae.

Furthermore, I will discuss only {\it circumstellar} material and not
{\it interstellar} material (i.e. H~{\sc ii} regions,
photodissociation regions, giant molecular clouds; e.g., Zinnecker,
these proceedings).  Similarly, I will not delve into circumstellar
accretion disks and outflows associated with the earliest formation
phases of massive stars, as these were also discussed earlier in this
conference (Beuther; these proceedings), nor will I discuss the
phenomena of pulsar winds or black hole accretion disks.  The main
focus here is the circumstellar matter that results from the mass loss
from massive stars in the course of their evolution up to core
collapse.

It is surprisingly difficult to directly detect circumstellar gas and
dust around extremely luminous central stars, and in part of this
contribution I will review some of the tricks observers use to see
circumstellar material in spite of the bright photospheric continuum
light.  In general, circumstellar nebulae can only be detected if the
material is extremely dense and located relatively far from the glare
of the star.  This requires densities several orders of magnitude
higher than the densities of stellar winds at the same radius.  Such
high densities, in turn, imply that the nebulae are created by large
amounts of matter ejected by the star.  That is the chief reason they
are of interest; namely, circumstellar nebulae provide a fossil record
of the most important mass-loss phases experienced by stars.

Mass loss plays a critical role in the evolution of massive stars
(e.g., Chiosi \& Maeder 1986; Maeder \& Meynet 2000; Meynet, these
proceedings), and profoundly impacts the eventual supernova explosion
(e.g., Woosley et al.\ 1993).  Whether the mass loss that leads to
Wolf-Rayet stars and stripped envelope supernovae (i.e. Types Ib and
Ic) is due to steady winds, eruptions, or binary Roche lobe overflow
(RLOF), and at exactly what initial masses this occurs, is still
debated.  In a recent paper, I have reviewed the connection between
mass loss of progenitor stars and the types of observed supernovae
more extensively (Smith et al.\ 2010).

Most importantly, the observational determination of the physical
parameters of circumstellar nebulae --- mass, composition, expansion
age and kinetic energy, as well as global geometry and detailed
structure --- provide critical constraints for some of the most
prodigious mass-loss phases of massive stars.  This is particularly
instructive for illuminating cases when the mass loss was strongly
enhanced for a short time, and is therefore rarely observed directly
(such as in brief giant LBV eruptions), or when the duration (and
hence the cumulative mass-loss budget) is not well known.  The
chemical abundances of this ejected material tell us about the recent
evolutionary phases of the star.

\section{Types of Stars with Circumstellar Nebulae}

As noted above, in order for a circumstellar nebula to be detectable,
it must be very dense, with densities far exceeding those of normal
stellar winds.  Extremely high densities at large radii from the star
can be achieved in two basic ways: (1) A dense shell can result from a
sudden eruption or explosion that ejects a large amount of material
from the star.  This may happen, for example in an LBV eruption or in
a red supergiant pulsation.  The ejection of extremely dense shells
may be accompanied by strong cooling to make dense clumps, or even
dust grain formation, both of which may enhance the ability to be
detected. (2) A dense shell nebula can result at the interface when a
faster wind sweeps into a slow dense wind.  In either case, the
presence of a nebula requires a substantial change in the mass-loss
behavior of the star on a relatively short timescale.  For this
reason, nebulae tend to be associated with stars in late transitional
phases of evolution off the main sequence or immediately before a
supernova.  We do not generally see circumstellar nebulae around main
sequence stars.\footnote{Incidentally, the fact that observable
  signatures of dense circumstellar material are absent around
  main-sequence O-type stars indicates that these stars quickly clear
  away all natal disk material associated with the star formation
  process.}  Some of the key types of stars that are normally
associated with substantial circumstellar nebulae are outlined below,
proceeding from cool to hot temperatures.  I will try to highlight a
demonstrative example of each.

\subsection{Red Supergiants}

Red supergiants have slow, dense, dusty winds.  How the winds are
driven from the stars is not completely understood, but it is likely
that a combination of pulsations and radiation force on dust grains is
at work, although there is also evidence for strongly enhanced
episodes of mass loss as in the case of extreme red supergiants like
VY CMa (Smith et al.\ 2001; 2009).  Circumstellar material can
sometimes be seen around red supergiants, but most notably in very
nearby or very extreme cases.  The detectability of circumstellar
material is enhanced by the fact tht the winds are slow and clumpy,
giving rise to high density regions.

Betelgeuse is an extremely nearby example of a relatively normal red
supergiant, but even at such close distances of only 150--200 pc
(Harper et al.\ 2008), its circumstellar material is difficult to
detect because its stellar wind mass-loss rate is only about 10$^{-6}$
$M_{\odot}$ yr$^{-1}$ (Harper et al.\ 2008; Smith et al.\ 2009).
Emission from its dusty wind has been resolved with mid-IR nulling
interferometry (Hinz et al.\ 1998), and its circumstellar shell has
been spatially resolved in emission lines like K~{\sc i} (Plez \&
Lambert 1994, 2002) and infrared CO bandhead emission (Smith et al.\
2009).

VY CMa is a much more striking case, where the recent mass-loss rate
is about 10$^3$ times stronger than Betelgeuse, and is thought to be
due to an enhanced mass-loss episode in the last 1,000 yr (Smith et
al.\ 2001, 2009).  This results in a dramatic circumstellar reflection
nebula that is easily detected in visual-wavelength images with {\it
  HST} (Smith et al.\ 2001), polarized light (Jones et al.\ 2007), IR
continuum emission from dust (Monnier et al.\ 1998), and in various
spectral lines like K~{\sc i} (Smith 2004), infrared CO bandhead
emission (Smith et al.\ 2009), and rotational lines of CO (Decin et
al.\ 2006).  Smith et al.\ (2009) have noted the stark difference
between Betelgeuse and VY CMa, using the same techniques to observe
circumstellar material around both stars.

\subsection{Yellow Hypergiants}

Like red supergiants, the yellow hypergiants (YHGs) have slow and
dense, dusty winds which can give rise to detectable nebulae.  These
cases are rare, however, and most YHGs do not have easily detectable
circumstellar nebulae.  One dramatic example of a YHG with an
observable nebula is IRC+10420 (see Oudmaijer 1998, Oudmaijer e al.\
1996; Humphreys et al.\ 1997, 2002; Davies et al.\ 2007), which seems
to be cruising across the top of the HR diagram, transitioning from a
spectral type of late F to early A in just a few decades.  If a YHG is
seen to have a spatially resolved nebula, it is thought to result
because the YHG is in a post-RSG phase.  In the case of IRC+10420 this
may be following a phase of enhanced RSG mass loss like VY CMa.
Otherwise it would be quite difficult to explain the presence of OH
masers (e.g., Bowers 1984) around such a warm star.  These nebulae are
dusty, seen in scattered starlight or thermal-IR emission, as well as
in molecular transitions at longer wavelengths (Tiffany et al.\ 2010;
Castro-Carrizio et al.\ 2001, 2007).

\subsection{Luminous Blue Variables}

LBVs are perhaps the best known examples of circumstellar nebulae
around massive stars, exemplified in memorable {\it HST} images like
those of $\eta$ Carinae (Morse et al.\ 1998) and the Pistol star
(Figer et al.\ 1999).  They are reminiscent of planetary nebulae in
their complex structure and geometry, although LBV nebulae can be much
more massive (see contributions by Vamratira-Nikov et al., Weis, Clark
et al., and Wachter in these proceedings).  Smith \& Owocki (2006)
noted several cases of luminous LBVs with nebulae of 10--20
$M_{\odot}$.  These extremely massive shells -- ejected relatively
recently (typical ages of roughly 10$^4$ yr or less) -- indicate an
extremely violent history of mass ejection, when these stars can
potentially shed a large fraction of their initial mass in a
disruptive event that lasts only a few years.  This eruptive mass loss
endures as one of the chief mysteries of stellar astrophysics, despite
its importance in determining the fate of massive stars.  Not all LBV
nebulae are so massive, of course.  Many are only of order 0.1
$M_{\odot}$.  This is sometimes even seen in the same star: after
ejecting $\sim$15 $M_{\odot}$ in its 1840s eruption, $\eta$ Carinae
subsequently ejected 0.1--0.2 $M_{\odot}$ in its smaller 1890 eruption
(Smith 2005).

The chemical abundances of LBV nebulae are generally nitrogen rich,
indicating that nuclear material processed through the CNO cycle has
risen to the surface of the star and has been ejected, therefore
indicating that these stars are in an advanced phase of their
evolution (Davidson et al.\ 1986; Lamers et al.\ 2001; Smith \& Morse
2004).  In the case of $\eta$ Car, this N enhancement is very recent,
occurring in jut the past few thousand years (Smith \& Morse 2004;
Smith et al.\ 2005).

The geometries of LBV nebulae are also interesting.  LBV shells are
often --- although not always --- bipolar.  An obvious extreme example
is $\eta$~Carinae, where the bipolar shape of the nebula seems
consistent with expectations of mass loss from a rapidly rotating star
(Smith 2006; Owocki 2003; Owocki et al.\ 1996; Dwarkadas \& Owocki
2002; Smith \& Townsend 2007).  Bipolarity of LBV nebulae has often
been attributed to various degrees of asymmetry in the pre-existing
ambient material (e.g., Frank et al.\ 1995), although the origin of
that pre-existing asymmetry does not have a clear explanation.  Recent
imaging of the nebula around the LBV HD~168625 (Smith 2007) showed a
triple-ring structure almost identical to the nebula around SN~1987A
(Burrows et al.\ 1995), providing a tantilizing link between LBVs and
supernova progenitors (see below). On the other hand, some LBVs are
only mildy ellipsoidal, like AG Car (see Weis, these proceedings), and
some like P Cygni appear to be clearly spherical (Smith \& Hartigan
2006).  This suggests that we should avoid impulses to associate {\it
  all} LBV eruptions with stellar mergers or similar binary-induced
effects.

\subsection{B[e] supergiants and Equatorial Rings}

The circumstellar material around B[e] stars was discussed in detail
by Zickgraf et al.\ (1986, 1996; see posters at this meeting by
Millour et al.; and Bonanos et al.).  B[e] stars are thought to be
surrounded by dusty equatorial tori with radii of order 10$^3$ AU,
making them bright IR sources, while gas in these slowly expanding
tori give rise to their relatively narrow namesake forbidden emission
lines.  In many ways, the B[e] supergiants resemble less extreme
versions of LBV nebulae. The spectral energy distributions of B[e]
supergiants in the SMC and LMC look very similar to those of LBVs
(Bonanos et al.\ 2009, 2010).

The origin of the equatorial circumstellar nebulae of B[e] stars
remain unclear, but possibilities are that they arise from post-RSG
evolution, equatorial mass loss from a recent RLOF phase, or that they
arise from rapidly rotating stars.  More detailed studies of B[e]
stars and their circumstellar matter are certainly justified.

The dusty tori around B[e] supergiants with radii of $\sim$10$^3$ AU
may be related to an emerging class of early B supergiants with
spatially resolved equatorial rings (see Smith et al.\ 2007; Smith
2007), including the progenitor of SN~1987A, SBW1 in the Carina
Nebula, Sher~25 in NGC~3603, and HD~168625.  Perhaps these equatorial
rings are the expanding fossil remains of the B[e] tori (Smith et al.\
2007).

\subsection{Interacting Binaries}

Although B[e] stars and LBVs are sometimes suspected to be binaries
(and are indeed sometimes known to be binaries where the role of the
companion star is unclear), there are also more clear-cut cases of
binary-induced mass loss.  Namely, there are very close interacting
binaries that are sometimes even seen as eclipsing binaries, where we
can see that they are in (or have recently been in) a RLOF phase of
evolution.

One of the most interesting cases to mention is the massive,
eclipsing, over-contact binary RY Scuti.  To my knowledge, it is so
far the only massive binary caught in the brief RLOF phase that also
has a spatially resolved circumstellar nebula.  This is interesting,
because the nebula is not only torroidal (like those around B[e]
supergiants), but also exhibits a bizarre double-ring structure (Smith
et al.\ 2002; Gehrz et al.\ 2001). The properties of the nebula were
discussed by Smith et al.\ (2002), while spectroscopy of the central
binary system has been discussed recently by Grundstrom et al.\
(2007).  RY Scuti is extremely interesting, since the originally more
massive member is thought to be caught in a brief transition to a WR
star.  This system therefore provides the rare opportunity to actually
watch the stripping of the H envelope and the origin of a Type Ibc
supernova progenitor in a binary system (e.g., Paczynski 1967).  Smith
et al.\ (2010) have concluded that this channel probably dominates the
production of Type~Ibc supernovae.

\subsection{Wolf-Rayet Stars}

Although Wolf-Rayet (WR) stars are famous for their strong winds, they
also sometimes exhibit circumstellar nebulae, which come in two main
flavors.  One is the large wind-blown bubble nebulae that are
generally seen around younger WN stars (see Stock \& Barlow 2010).  A
famous example is NGC~6888.  The raw material for these nebulae
probably is not from the WR wind itself, however.  Rather, the dense
nebular gas is probably slower material ejected in a previous LBV or
RSG phase, which is then swept into a dense bubble or shell by the
faster WR wind.

A very different type of circumstellar nebula seen around WR stars is
the dusty nebulae associated with WC+O binaries, either appearing as
so-called ``pinwheel'' nebulae in circularized systems like WR104, or
as episodic puffs of asymmetric dust production in eccentric systems
like WR140.  Although the dusty nature of WC stars had been known
since the early days of IR astronomy (Gehrz \& Hackwell 1974), the
spectacular structure of these WR nebulae was only revealed by special
high-resolution aperture synthesis imaging at near-IR wavelengths
pioneered on the Keck telescope (see Tuthill et al.\ 1999; Monnier et
al.\ 1999, 2002). In these cases, the dense gas and dust -- which is
much denser than the material normally found in a WR star wind --
arises as a result of the compression and cooling in the
colliding-wind shock of the WC+O binary (see Williams et al.\ 1990,
2001).  The formation of graphite grains is likely facilitated by the
C-rich material in the WC wind.


\section{Multiwavelength Diagnostics of Circumstellar Material}

\subsection{Visual Wavelengths}

At visual wavelengths, the two chief ways to spatially resolve
circumstellar material are with starlight scattered by dust, and with
intrinsic emission lines in the nebula.  Because the central stars are
bright, simple optical continuum imaging is only successful in a few
remarkable cases, like $\eta$~Carinae, VY~CMa, IRC+10420, etc. (see
above), where the dusty nebula is very dense and the central star is
partly obscured by the circumstellar dust.  In other cases,
polarimetric imaging or coronagraphy can help to suppress the direct
light from the central star.

A more effective method is to use narrow-band imaging, long-slit
spectroscopy, or IFU spectroscopy to detect extended emission lines
from the circumstellar nebula. Concentrating on a nebular emission
line formed only in the nebula helps suppress the bright continuum
radiation from the central star.  The most common probe is [N~{\sc
  ii}] $\lambda$6583, which arises in most circumstellar nebulae over
a wide range in ionization level, and can be especilly bright in the
N-enriched gas around evolved massive stars (Davidson et al.\ 1986;
Lamers et al.\ 2001; Smith \& Morse 2004).  Forbidden lines like
[N~{\sc ii}] are usually better than H$\alpha$, since H$\alpha$ is
often an extremely strong emission line in the central star's spectrum
as well.  When high-resolution long-slit spectra are employed, lines
like [N~{\sc ii}] permit one to measure the expansion speed of a
circumstellar shell (e.g., L.\ Smith 1994).

A lesser-known technique that has proven extremely useful for
detecting nebulae around cooler stars (where N is neutral and [N~{\sc
  ii}] cannot be seen) is the red resonance lines of K~{\sc i}.
Extended K~{\sc i} $\lambda$7699 emission has been studied in detail
around Betelgeuse and VY~CMa, for example (Bernat \& Lambert 1976;
Bernet et al.\ 1978; Plez \& Lambert 1994, 2002; Smith 2004).  With
high-resolution long-slit spectroscopy of K~{\sc i} $\lambda$7699, one
can perform the same types of kinematic studies as with [N~{\sc ii}]
around hotter stars.  For nebulae around very hot stars, higher
ionization lines like [Fe~{\sc iii}] can also be useful, as in the
case of RY Scuti (e.g., Smith et al.\ 2002).

\begin{figure}[h]
\centering
\includegraphics[width=15cm]{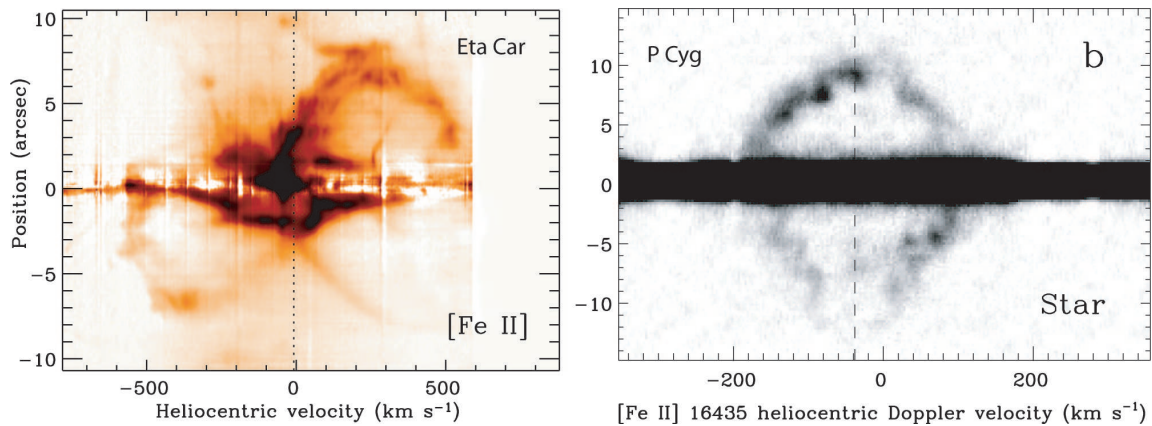}
\caption{Long slit spectra of [Fe II] 1.644 $\mu$m around Eta Carinae
  (left; a) and P Cygni (right; b).  The spectrum of Eta Car was taken
  with Phoenix on Gemini South (Smith 2006).  The spectrum of P Cygni
  was taken with CSHELL on the IRTF (Smith \& Hartigan 2006).}
\end{figure}

\subsection{Near-IR}

Near-IR wavelengths can provide a powerful probe of circumstellar
material, especially in cases of massive stars that are obscured at
visual wavelengths (in the Galactic Center, for example; Figer et al.\
1999; Mauerhan et al.\ 2010).  As with optical imaging, circumstellar
shells can be resolved using scattered continuum starlight, although
this is perhaps even more difficult than at visual wavelengths because
of the lower scattering efficiency, unless the grains are
large. However, with available technology, near-IR wavelengths have
the advantage of adaptive-optics (AO) imaging or speckle-masking
interferometry, as mentioned earlier.  Again, polarimetric imaging and
coronagraphy can help to suppress the bright central starlight.

Narrow-band imaging and long-slit spectroscopy are also used to
resolve circumstellar material around massive stars in the near-IR,
with H~{\sc i} lines such as Br~$\gamma$ or (from space with {\it
  HST}) Pa$\alpha$ (e.g., Figer et al.\ 1999; Mauerhan et al.\ 2010).
These lines are best around hotter stars where H is mostly ionized.
In principle this is the same as H$\alpha$ imaging in the optical, but
it can be used for obscured sources in the Galactic plane.

One of the most powerful but underused near-IR probes of circumstellar
gas around massive stars involves spectroscopy or narrow-band imaging
of infrared [Fe~{\sc ii}] emission lines.  In particular, [Fe~{\sc
  ii}] $\lambda$16435 and $\lambda$12567 are two of the brightest
lines in the near-IR spectra of LBVs and similar stars (Smith 2002),
due to the relatively low excitation and high density of the gas.
These two bright lines arise from the same upper energy level, and so
their observed ratio can serve as a reliable measure of the reddening
and extinction toward a source (see Smith \& Hartigan 2006 for the
atomic data and intrinsic line ratios).  Flux ratios of some other
adjacent [Fe~{\sc ii}] lines to [Fe~{\sc ii}] $\lambda$16435 serve as
density diagnostics, while high-resolution spectra can provide the
expansion speed of a shell.  Detailed studies of $\eta$ Car and
P~Cygni have demonstrated the utility of these [Fe~{\sc ii}] lines
(Smith 2006; Smith \& Hartigan 2006; see Figure 1).

Narrow-band imaging or spectroscopy of near-IR H$_2$ lines, like H$_2$
1--0 S(1) at 2.122 $\mu$m, are also common tracers of shocked gas or
dense gas irradiated by moderately strong non-ionizing radiation.
However, these lines are rarely seen around hot stars because the
H$_2$ is destroyed, and they are rarely seen in cooler supergiants
because they are not sufficiently excited.  An unusual exception is
the extremely bright H$_2$ lines in the Homunculus of $\eta$~Carinae
(Smith 2006).  In this source, the combination of a very young and
dense nebula that is optically thick enough to be self-shielding
allows the H$_2$ molecules to survive, while a surface layer of H$_2$
is struck by strong near-UV radiation from the luminous central star.
As the Homunculus continues to expand and become more optically thin,
the H$_2$ will be dissociated (see Smith \& Ferland 2006 for details).

\begin{figure}[h]
\centering
\includegraphics[width=15cm]{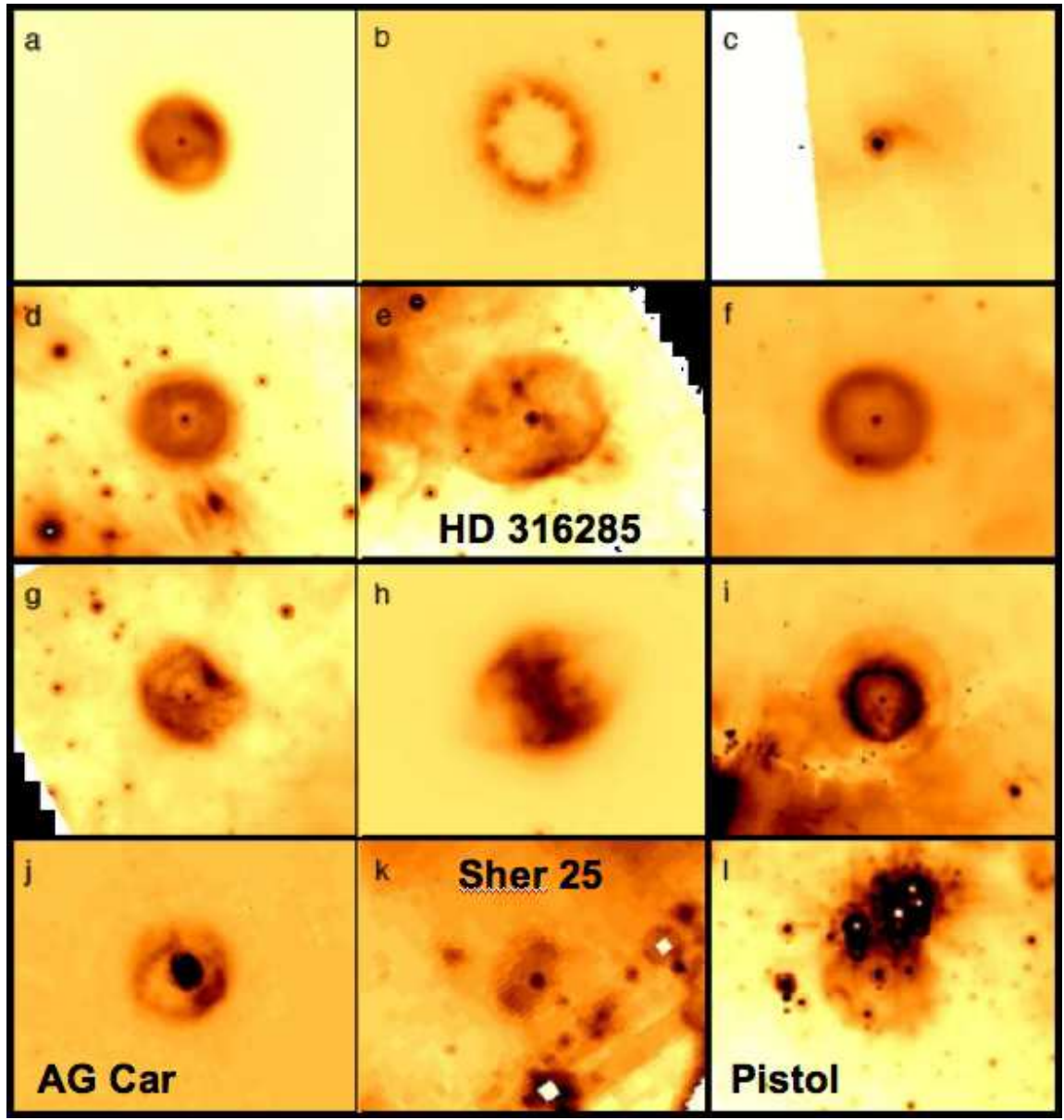}
\caption{{\it Spitzer}/MIPS 24~$\mu$m images of shells around LBVs and
  related stars (adapted from Gvaramadze et al.\ 2010).}
\end{figure}

\subsection{Mid-IR/Far-IR}

Moving into the mid-IR and far-IR, circumstellar shells of massive
stars become more easily observed because the glare of photospheric
emission from the central stars is no longer a problem, and because
thermal-IR radiation from warm dust grains is substantial.  Extreme
red supergiants like VY~CMa and NML~Cygni have been favorite targets
of ground-based mid-IR observers for decades because their
circumstellar dust is bright and spatially extended, while
$\eta$~Carinae is observed perpetually with every new mid-IR
instrument that comes online in the southern hemisphere.  Mid-IR
observations played a special role in the history of our understanding
of $\eta$~Carinae, providing our first clue that it is an extremely
luminous but self-obscured star (Westphal \& Neugebauer 1969).  The
young dust shell of $\eta$~Carinae acts as a calorimeter of the central
star because it absorbs nearly all the star's UV radiation, and
mid/far-IR observations provide our best estimates of the mass of the
Homunculus (Smith et al.\ 2003).

Eta Carinae is an exceptional case, however, because its nebula is
very young and very bright.  For other LBVs, the shells are not so
easy to detect in ground-based images because they are larger and more
optically thin, and are therefore often too faint to detect in the
mid-IR through the Earth's atmosphere.  However, space-based IR
telescopes have provided key information on a number of dusty shells
around LBVs (e.g., Voors et al. 2000; Trams et al.\ 1998; Egan et al.\
2002; Clark et al.\ 2003).  In particular, recent surveys of the
Galactic plane with the IRAC and MIPS instruments on {\it Spitzer}
have revealed a large number of extended mid-IR shells around LBVs and
related stars (see Fig.\ 2; Gvardmadze et al.\ 2010; Wachter et al.\
2010; Smith 2007).  This sample has the potential to tell us how much
mass is typically ejected by an LBV, and to identify previously
unrecognized LBVs and WR stars.  We eagerly anticipate results on the
far-IR emission from these shells with {\it Herchel}.  Longer mid-IR
and far-IR wavelengths are useful, because they have the potential to
detect cooler dust in the shell, which may correspond to a large
fraction of the total mass.

In addition to very extended shell nebulae, the more compact dusty
pinwheels and episodic ejections associated with colliding-wind WC+O
binaries are also spatially resolved in the thermal-IR, as noted
earlier (Tuthill et al.\ 1999; Monnier et al.\ 1999, 2002), providing
a powerful probe of a unique mass-loss phenomenon.

\subsection{X-rays}

The spatially resolved nebulae around massive stars are not often
detected in X-rays, since strong shocks (and hence, strong differences
in ejection velocity over a short period of time) are needed to
produce sufficiently bright X-rays far from the star.  The study of
{\it diffuse} X-ray emission asociated with massive stars is mainly
concentrated toward SNe and SN remnants (see below).  A notable
exception, again, is the peculiar case of $\eta$~Carinae, where a
strong blast wave from the 19th century eruption is overtaking ejecta
from a previous eruption (Smith 2008), giving rise to a spectacular
soft X-ray shell made famous in {\it Chandra} images (e.g., Seward et
al.\ 2001).  The study of massive stars in X-rays is weighted heavily
toward colliding wind binaries or unresolved soft X-ray emission from
the winds of O-type stars.

\subsection{Radio}

At radio wavelengths, continuum free-free radiation traces the same
photoionized gas around hot massive stars that can be observed with
H$\alpha$ emission, but without such strong continuum radiation from
the central star, and free from line-of-sight extinction.  This is
particularly useful for studying the nebulae of hotter LBVs and WR
stars (e.g., Duncan \& White 2002).

Radio wavelengths also provide unique probes of molecular shells
around massive stars, most commonly seen around cooler supergiants.
In particular, molecular masers like SiO, H$_2$O, and OH can be
observed at very high spatial resolution with radio interferometers,
and have yielded unique and valuable information about the structure
and expansion of shells around evolved cool stars (e.g., Bowers et
al.\ 1993; Benson \& Mutel 1979, 1982; Marvel 1997; Boboltz \& Marvel
2000; Trigilio et al.\ 1998).  One can actually follow the proper
motion and Doppler velocity of individual maser spots, tracing out the
structure, expansion, and rotation of the inner winds of these stars.
Molecular shells are not generally observed around hot stars because
they are quickly dissociated by UV radiation.  However, they are seen
in some cases of young WR stars or YHGs that are in a post-RSG phase,
as noted earlier for IRC+10420.  The N-enriched shells of LBVs may be
detectable in ammonia, which has been detected in $\eta$ Car (Smith et
al.\ 2006), but has not been searched for extensively in other
sources.

\section{Supernova Blast Waves Crashing into Pre-Supernova
  Circumstellar Material}

When a supernova (SN) explodes, it sends a flash of UV radiation and a
strong blast wave out into the surrounding medium.  In this way, SNe
illuminate the circumstellar material that was ejected by the star
{\it before} the SN.  As the shock expands it can interact with either
circumstellar or interstellar material, giving rise to a SN remnant.
The posterchild for a SN remnant interacting with dense circumstellar
material ejected by its progenitor is Cas~A (e.g., Chevalier \& Oishi
2003).  Slow-moving N-rich ``flocculi'' (Fesen \& Becker 1991;
Chevalier \& Kirshner 1978) indicate that the overtaken material was
circumstellar rather than interstellar, and that the progenitor was at
least a moderately massive evolved star with a fairly slow wind
(Chevalier \& Oishi 2003).

Perhaps the most famous case of a SN blast wave interacting with
circumstellar material is SN~1987A.  Almost immediately, the UV flash
of the SN shock breakout photoionized the triple ring nebula seen in
{\it HST} images (Burrows et al.\ 1995).  After about 10 years, the SN
blast wave began crashing into the equatorial ring seen in these
images, and this collision is still unfolding (Michael et al.\ 2000;
Sonneborn et al.\ 1997; Sugerman et al.\ 2002).  The complex shock
interaction is the focus of an ongoing multiwavelength campaign, and
provides an enormous reservoir of information about shock physics as
well as pre-SN mass loss of the progenitor star.  Since the blue
progenitor contradicted expectations of stellar evolution models, it
has been suggested that the progenitor star underwent a binary merger
event to produce the triple-ring nebula (e.g., Morris \& Podsiadlowski
2007), but comparisons to LBV nebulae have also been made (Smith
2007).

\begin{figure}[h]
\centering
\includegraphics[width=12cm]{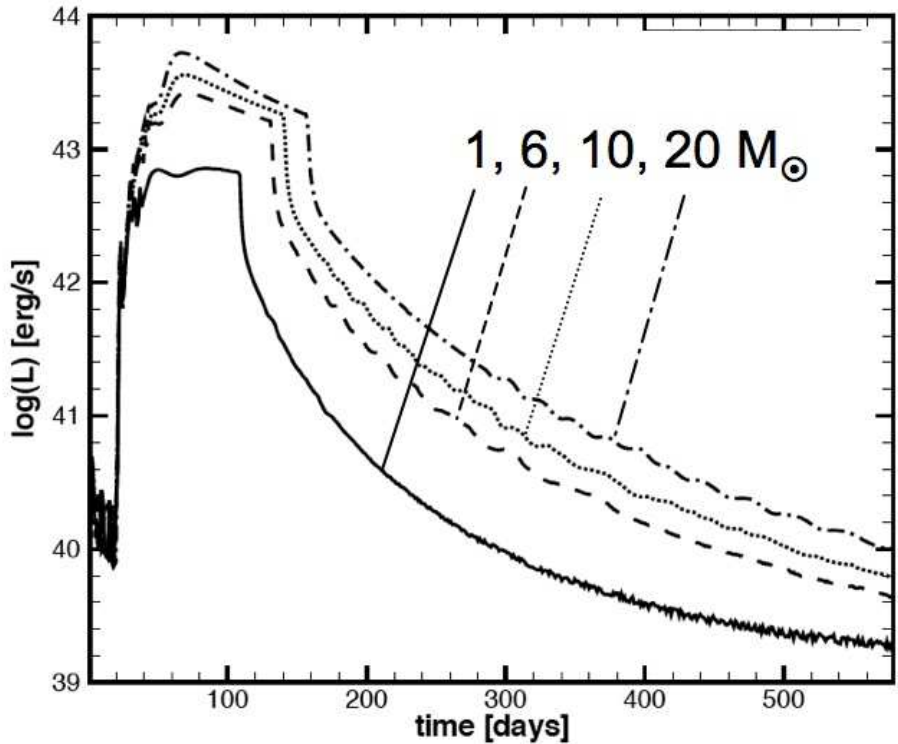}
\caption{Theoretical light curves of a normal SN shock crashing into
  dense circumstellar gas with a range of masses from 1--20
  $M_{\odot}$ (from van Marle et al.\ 2010). A normal Type II-P
  supernova has a peak luminosity of about 10$^{42}$ erg s$^{-1}$.}
\end{figure}

Sometimes the interaction between the SN shock and circumstellar gas
happens much sooner and is even more dramatic.  In the case of
Type~IIn, this shock interaction with the nearest circumstellar gas
can happen immediately, completely altering the apparent spectrum of
the SN and in some cases markedly increasing the luminosity of the SN.
The name ``IIn'' comes from the narrow H emission lines in the
spectrum, arising from the slow shocked CSM gas.  The dense
circumstellar material decelerates the blast wave and thermalizes its
kinetic energy.  In some cases this thermal energy is radiated as
visual light before adiabatic expansion can cool the gas, thus
producing some of the most luminous SNe in the universe with
$\sim$10$^{51}$ ergs radiated in visual light alone (e.g., Smith \&
McCray 2007, Smith et al.\ 2007, 2010; Woosley et al.\ 2007; van Marle
et al.\ 2010; see Figure 3).  By studying the time evolution of the
luminosity and spectral properties of SNe~IIn in the year or two after
discovery, we can deduce the density and velocity of circumstellar
matter at each radial position overtaken by the shock.  An exemplary
case concerns the spectral evolution and light echo of SN~2006gy, from
which a circumstellar medium closely resembling that of $\eta$~Carinae
has been deduced (Smith et al.\ 2010).  We can therefore reconstruct
the mass-loss rate and kinetic energy of pre-SN mass ejections as a
function of time immediately before the SN explosion occurred.  Aside
from being lucky enough to be watching when an exceptional Galactic SN
occurs, this is our most powerful tool for studying how massive stars
behave in the rapid nuclear timescales immediately before they suffer
core collapse.

The potential of using SNe~IIn to learn about circumstellar structure
and pre-SN evolution is extremely exciting.  Not only does this shock
interaction produce some of the most luminous SNe known, potentialy
observable at high redshift, but it also can provide diagnostics of
the detailed properties of circumstellar gas around a single (or
binary) star at distances far beyond where we could ever hope to
spatially resolve a circumstellar nebula.

\section*{Acknowledgements}

I thank the conference organizers for covering part of my travel
expenses at the conference.

\footnotesize
\beginrefer

\refer Benson, J.M., \& Mutel, R.L.\ 1979, ApJ, 233, 119

\refer Benson, J.M., \& Mutel, R.L.\ 1982, ApJ, 253, 199

\refer Bernet, A.P., \& Lambert, D.L.\ 1976, ApJ, 210, 395

\refer Bernet, A.P., et al.\ 1978, ApJ, 219, 532

\refer Boboltz, D.A., \& Marvel, K.B.\ 2000, ApJ, 545, L149

\refer Bonanos, A., et al. 2009, AJ, 138, 1003

\refer Bonanos, A., et al.\ 2010, AJ, 140, 416

\refer Bowers, P.F.\ 1984, 279, 350

\refer Bowers, P.F., Claussen, M.J., \& Johnston, K.J.\ 1993, AJ, 104,
284

\refer Burrows, C.J., et al.\ 1995, ApJ, 452, 680

\refer Castro-Carrizo, A., Lucas, R., Bujarrabal, V., Colomer, F., \&
Alcolea, J.\ 2001, A\&A, 368, L34

\refer Castro-Carrizo, A., Quintana-Lacaci, G., Bujarrabal, V., Neri,
R., \& Alcolea, J.\ 2007, A\&A, 465, 457

\refer Chevalier, R.A., \& Kirshner, R.P.\ 1978, ApJ, 219, 931

\refer Chevalier, R.A., \& Oishi, J.\ 2003, ApJ, 593, L23

\refer Chiosi, C., \& Maeder, A.\ 1986, ARAA, 24, 329

\refer Clark, J.S., Egan, M., Crowther, P., et al.\ 2003, A\&A, 412, 185

\refer Davidson, K., Dufour, R., Walborn, N.R., \& Gull, T.R.\ 1986,
ApJ, 305, 867

\refer Davies, B., Oudmaijer, R.D., \& Sahu, K.C.\ 2007, ApJ, 671,
2059

\refer Decin, L., et al.\ 2006, A\&A, 456, 549

\refer Duncan, R.A., \& White, S.M.\ 2002, MNRAS, 330, 63

\refer Dwarkadas, V.V., \& Owocki, S.P.\ 2002, ApJ, 581, 1337

\refer Egan, M., Clark, J.S., Mizuno, D.R., et al.\ 2002ApJ, 572, 288

\refer Fesen, R.A., \& Becker, R.H.\ 1991, ApJ, 371, 621

\refer Figer, D., McClean, I.S., \& Morris, M.\ 1999, ApJ, 514, 202

\refer Frank, A., Balick, B., \& Davidson, K.\ 1995, ApJ, 441, L77

\refer Gehrz, R.D., \& Hackwell, J.A.\ 1974, ApJ, 194, 619

\refer Gehrz, R.D., Smith, N., \& Jones, B., et al.\ 2001, ApJ, 559, 395

\refer Grundstrom, E., et al.\ 2007, ApJ, 667, 505

\refer Gvaramadze, V., et al.\ 2010, MNRAS, 405, 1047

\refer Harper, G., Brown, A., \& Guinan, E.\ 2008, AJ, 135, 1430

\refer Hinz, P.M., et al.\ 1998, Nature, 395, 251

\refer Humphreys, R.M., Smith, N., Davidson, K., et al.\ 1997, AJ,
114, 2778

\refer Humphreys, R.M., Davidson, K., \& Smith, N.\ 2002, AJ, 124,
1026

\refer Jones, T.J., et al.\ 2007, AJ, 133, 2730

\refer Lamers, H.J.G.L.M., et al.\ 2001, ApJ, 551, 764

\refer Maeder, A., \& Meynet, G.\ 2000, A\&A, 361, 159

\refer Marvel, K.B.\ 1997, PASP, 104, 1286

\refer Mauerhan, J., et al.\ 2010, arXiv:1009.2769

\refer Michael, E., et al.\ 2000, ApJ, 542, L53

\refer Monnier, J.D., Geballe, T.R., \& Danchi, W.C.\ 1998, ApJ, 502,
833

\refer Monnier, J.D., Tuthill, P., \& Danchi, W.C.\ 1999, ApJ, 525, L97

\refer Monnier, J.D., Tuthill, P., \& Danchi, W.C.\ 2002, ApJ, 567, L137

\refer Morris, T., \& Podsiadlowski, P.\ 2007, Science, 351, 1130

\refer Morse, J.A., et al.\ 1998, AJ, 116, 2443

\refer Oudmaijer, R.D.\ 1996, A\&AS, 129, 541

\refer Oudmaijer, R.D., Groenewegen, M.A.T., Matthews, H.E.,
Blommaert, J.A.D.L., \& Sahu, K.C.\ 1996, MNRAS, 280, 1062

\refer Owocki, S.P.\ 2003, in IAU Symp. 212, 281

\refer Owocki, S.P., Cranmer, S.R., \& Gayley, K.G.\ 1996, ApJ, 472,
L115

\refer Plez, B., \& Lambert, D.L.\ 1994, ApJ, 425, L101

\refer Plez, B., \& Lambert, D.L.\ 2002, A\&A,386, 1009

\refer Paczy´nski, B. 1967, Acta Astron., 17, 355

\refer Seward, F., et al.\ 2001, ApJ, 553, 832

\refer Smith, L.\ 1994, ApSS, 216, 291 

\refer Smith, N.\ 2002, MNRAS, 336, L22

\refer Smith, N.\ 2004, MNRAS, 349, L31

\refer Smith, N.\ 2005, MNRAS, 357, 1330

\refer Smith, N.\ 2006, ApJ, 644, 1151

\refer Smith, N.\ 2007, AJ, 133, 1034

\refer Smith, N.\ 2008, Nature, 455, 201

\refer Smith, N., \& Ferland, G.R.\ 2007, ApJ, 655, 911

\refer Smith, N., \& Hartigan, P.\ 2006, ApJ, 638, 1045

\refer Smith, N., \& McCray, R.\ 2007

\refer Smith, N., \& Morse, J.A.\ 2004, ApJ, 605, 854

\refer Smith, N., \& Owocki, S.P.\ 2006, ApJ, 645, L45

\refer Smith, N., \& Townsend, R.H.D.\ 2007, ApJ, 666, 967

\refer Smith, N., et al.\ 2001, AJ, 121, 1111

\refer Smith, N., Gehrz, R.D., Hinz, P.M., et al.\ 2003, AJ, 125, 1458

\refer Smith, N., et al.\ 2002, ApJ, 578, 464


\refer Smith, N., Brooks, K.J., Koribalski, B.S., \& Bally, J.\ 2006,
ApJ, 645, L41

\refer Smith, N., Morse, J.A., \& Bally, J.\ 2005, AJ, 130, 1778

\refer Smith, N., Bally, J., \& Walawender, J.\ 2007, AJ, 134, 846

\refer Smith, N., Hinkle, K.H., \& Ryde, N.\ 2009, AJ, 137, 3558

\refer Smith, N., et al.\ 2010, ApJ, 709, 856

\refer Smith, N., et al.\ 2010, MNRAS, in press; arXiv:1006.3899

\refer Sonneborn, G., et al.\ 1998, ApJ, 492, L139

\refer Stahl, O.\ 1987, A\&A, 182, 229

\refer Sugerman, B.E.K.\ 2000, ApJ, 572, 209

\refer Tiffany, C., Humphreys, R.M., Jones, T.J., \& Davidson, K.\
2010, AJ, 140, 339

\refer Trams, N.R., et al.\ 1998, Ap\&SS, 255, 195

\refer Trigilio, C., Umana, G., \& Cohen, R.J.\ 1998, MNRAS, 297, 497

\refer Tuthill, P.C., Monnier, J.D., \& Danchi, W.C.\ 1999, Nature,
398, 486

\refer Voors, R.H.M., et al.\ 2000, A\&A, 356, 501

\refer Wachter, S., et al.\ 2010, AJ, 139, 2330

\refer Westphal, J.A., \& Neugebauer, G.\ 1969, ApJ, 156, L45

\refer Williams, P.M., et al.\ 1990, MNRAS, 243, 662

\refer Williams, P.M., et al.\ 2001, MNRAS, 324, 156

\refer Woosley, S.E., Langer, N., \& Weaver, T.A.\ 1993, ApJ, 411, 823

\refer Zickgraf, F.J., et al.\ 1986, A\&A, 163, 119

\refer Zickgraf, F.J., et al.\ 1996, A\&A, 309, 505

\endrefer           
\end{document}